\documentclass[12pt]{article}

\begin{document}

\centerline{HAMILTON'S ECCENTRICITY VECTOR} 
\centerline{GENERALISED TO NEWTON WONDERS}

\bigskip

\centerline{\it By D. Lynden-Bell}
\centerline{\it Institute of Astronomy}

\bigskip

\large
The vectorial velocity is given as a function of the position of a
particle in orbit when a Newtonian central force is supplemented by an
inverse cubic force as in Newton's theorem on revolving orbits.

Such expressions are useful in fitting orbits to radial velocities of
orbital streams. The Hamilton-Laplace-Runge-Lenz eccentricity vector
is generalised to give a constant of the motion for these systems and
an approximate constant for orbits in general central potentials. A
related vector is found for Hooke's centred ellipse.

\normalsize
\bigskip

\noindent{\it Introduction}
\medskip

In a central orbit a particle with position $(r,\tilde{\phi})$ has
$r^2d\tilde{\phi}/dt=\tilde{h}$ constant.  Newton$^1$ pointed out that
if $\phi=\tilde{\phi}/n$, with $n$ any constant, then $r^2d\phi/dt=h$
is also constant. He then enquired what extra radial force would be
needed to make a new orbit with $\phi$ replacing $\tilde{\phi}$ but
with the $r(t)$ unchanged. Evidently $h=\tilde{h}/n$ so the old radial
equation of motion $d^2r/dt^2 -\tilde{h}^2 r^{-3}=\tilde{F}$ will
change to $d^2r/dt^2-h^2r^{-3}=F$.

Thus the extra radial force required is
$F-\tilde{F}=-(n^{-2}-1)\tilde{h}^2/r^3$ which is outward or inward
according as $n$ is greater or less than one. If the new motion were
observed from axes that rotate (non-uniformly) at the rate
$\Omega=(n^{-1}-1)\dot{\tilde{\phi}}$ then one would see the particle
perform the original orbit unchanged. Thus from fixed axes the whole
orbit may be thought of as revolving with angular velocity
$(n^{-1}-1)\dot{\tilde{\phi}}=(1-n)\dot{\phi}$. This is Newton's
theorem on revolving orbits. Newton used it to demonstrate the
accuracy of the inverse square law in the solar system; notice however
that the theorem holds for orbits subject to any central force
$\tilde{F}$. Chandrasekhar$^2$ gives an elegant discussion but see
Lynden-Bell \& Lynden-Bell$^3$ for the true shapes of Newton's orbits
in uniformly rotating axes.

Hereafter we specialise to an inverse square law supplemented by an
inverse cubic force so the potentials considered take the form
$$\psi=\mu r^{-1}+\frac{1}{2}Kr^{-2}$$
where $\mu$ may be thought of as $GM$ and $K$ is a constant which in
galactic orbits is normally negative though in Hartree Fock atoms it
is positive. The equation of motion in fixed axes is
$$d^2{\bf r}/dt^2=-(\mu r^{-2}+K r^{-3})\hat{\bf r}$$
where hats denote unit vectors.

Comparison with Newton's theorem on revolving orbits suggests that we
think of the angular momentum ${\bf h}$ as $n^{-1}\tilde{\bf h}$ and
$K$ as $(n^{-2}-1)\tilde{h}^2=(1-n^2)h^2$. If we then go to axes
rotating non-uniformly with angular velocity $(n^{-1}-1)\tilde{\bf
h}/r^2$, the orbital equations relative to the rotating axes must
reduce to those under the action of the inverse square law alone. In
practice it is $h$ and $K$ that are known, so in terms of those
$n^2=1-Kh^{-2}$. Should $K$ be greater than $h^2$ we would be in
trouble, but under such circumstances the inverse cube attraction
overcomes the centrifugal repulsion so orbits spiral into the
origin. Otherwise in the non-uniformly rotating axes with ${\bf
\Omega} = (1-\sqrt{1-Kh^{-2}}){\bf h}/r^2$ we recover writing a dot
for time derivatives relative to the rotating axes
\begin{equation} \label{1}
\ddot{\bf r}=-\mu r^{-2}\hat{\bf r} ~.
\end{equation}

We have indeed laboriously checked that with this ${\bf \Omega}$ the 2
${\bf \Omega}\times\dot{\bf r}+\dot{\bf \Omega} \times {\bf r} + {\bf
\Omega} \times ({\bf \Omega} \times {\bf r})$ terms cancel out with
the $-K\hat{\bf r}/r^3$ force term.

\bigskip
\noindent{\it The orbit seen in the rotating axes}
\medskip

From (\ref{1}) 
$${\bf r}\times\dot{\bf r}=\tilde{\bf h}$$
a constant, and 
$$\ddot{\bf r} \times \tilde{\bf h}=-\mu\hat{\bf r}\times
\left({\hat{\bf r}}\times \frac{\dot{\bf r}}{r}\right)=\mu\dot{\hat{\bf
r}}~,$$
thus
\begin{equation} \label{2}
\dot{\bf r}\times \tilde{\bf h}=\mu(\hat{\bf r} + \tilde{\bf e})~,
\end{equation}
where $\tilde{\bf e}$ is a constant vector \underline{fixed in these
rotating axes}. Since both the other terms are perpendicular to
$\tilde{\bf h}$, $\tilde{\bf e}$ must be too, so it lies in the plane of
the orbit. Dotting with $\hat{\bf r}/\mu$ we find setting
$l=\tilde{h}^2/\mu$,
\begin{equation} \label{3}
l/r=(1+\tilde{\bf e}. \hat{\bf r})=1+\tilde{e} \cos \tilde{\phi}~.
\end{equation}
This is the equation of a conic of eccentricity $\tilde{\bf e}$ and
semi-latus-rectum $l$ and $\tilde{\phi}$ is the angle in the rotating
axes measured from perihelion. Thus $\tilde{\bf e}$ points toward
perihelion in the rotating axes and it will inevitably rotate relative
to fixed axes. Since the axes rotate at the rate
$(n^{-1}-1)\tilde{h}/r^2=(1-n)\dot{\phi}$ we find
\begin{equation} \label{4}
\tilde{\bf e}={\bf e} \cos [(1-n)\phi]+\hat{\bf h}\times {\bf e} \sin
[(1-n)\phi] 
\end{equation}
where ${\bf e}$ is in the fixed absolute direction to the perihelion
at which $\phi=0$ and its magnitude is the eccentricity $|\tilde{\bf
e}|$. For fixed axes this comes from Hamilton$^4$. For the
contributions of Laplace, Runge \& Lenz see Goldstein$^5$.

\bigskip
\noindent{\it The Orbit seen from fixed axes}
\medskip

The equation of the orbit (\ref{3}) is put into fixed axes by writing
$\tilde{\phi}=n\phi$, $l=n^2h^2/\mu$
\begin{equation} \label{5}
l/r=1+e \cos n\phi~.
\end{equation}

The velocity in fixed axes will be 
$${\bf v} =\dot{\bf r}+{\bf \Omega} \times {\bf r}$$
where from (\ref{2})
$$\dot{\bf r}=\frac{\mu}{nh}\hat{\bf h}\times(\hat{\bf r}+\tilde{\bf
e})~.$$

Thus using (\ref{4}) for $\tilde{\bf e}$ and ${\bf
\Omega}=(1-n)\dot{\phi}\hat{\bf h}=(1-n){\bf h}/r^2$ we find our
expression for ${\bf v}$
\begin{equation} \label{6}
{\bf v}=\frac{\mu}{nh}\left\{\hat{\bf h}\times \hat{\bf r}\left[1+
\frac{l}{r}(n^{-1}-1)\right] +\hat{\bf h} \times {\bf e} \cos
[(1-n)\phi] -{\bf e} \sin [(1-n)\phi]\right\} 
\end{equation}
or using (\ref{5}) to eliminate $l/r$ in favour of $\phi$
\begin{equation} \label{7}
{\bf v} = \frac{\mu}{nh} \left\{\hat{\bf h}\times \hat{\bf r}\left[
n^{-1}+(n^{-1}-1) e \cos (n\phi )\right] + \hat{\bf h} \times {\bf e}
\cos [(1-n)\phi] - {\bf e} \sin [(1-n)\phi]\right\}~.
\end{equation}

To get the radial velocity along any line of sight $\hat{\bf l}$ one
merely uses $\hat{\bf l}.{\bf v}$ and then corrects for the Sun's
motion.

\bigskip
\noindent{\it The eccentricity vector constant of the motion}
\medskip

Equations (\ref{6}) or (\ref{7}) give us the velocity in terms of a
conserved constant vector ${\bf e}$ pointing toward the initial
perihelion. To get ${\bf e}$ itself we need to invert this equation so
${\bf e}$ is expressed as a function of ${\bf v}$ etc. To do this we
note that the final two terms in the $\{\}$ in (\ref{6}) yield ${\bf
e}$ if we first cross multiply them by $\times \hat{\bf h} \cos
[(1-n)\phi]$ and add the result to the same two terms $\times
(-\sin[(1-n)\phi])$. But from (\ref{6}) those final two terms are
equal to 
$$\frac{nh}{\mu} {\bf v} + \hat{\bf r}\times\hat{\bf h}\left[1+
\frac{l}{r} (n^{-1}-1)\right]$$
hence both $e^2$ is the square of this, which yields
$e^2=1+2l\mu^{-1}\varepsilon$ where
$\varepsilon=\frac{v^2}{2}-\frac{\mu}{r}-\frac{K}{2r^2}$ and
\begin{equation} \label{8}
{\bf e}\!=\!
\left\{\!\frac{nh}{\mu} {\bf v}\!\times\!\hat{\bf h}\!-\!\hat{\bf
r}\!\left[1\!+\!\frac{l}{r}(n^{-1}\!-\!1) \right]\!\right\} 
\cos[(1-n)\phi] -
\left\{\!\frac{nh}{\mu} {\bf v}\!+\!\hat{\bf r}\!\times\!\hat{\bf
h}\!\left[ 1\!+\!\frac{l}{r} (n^{-1}\!-\!1)\!\right]\!\right\} 
\sin[(1-n)\phi]
\end{equation}
which gives the new vector constant of the motion. We remind the
reader that $\phi=\cos^{-1}(\hat{\bf e}.\hat{\bf r})$ and
$n=\sqrt{1-Kh^{-2}}$. We may eliminate $l/r$ in terms of $\phi$ and
$e$ by using (\ref{5}). Notice that (\ref{8}) is actually an implicit
equation because $\phi$ is not known until $\hat{\bf e}$ is
known. However by taking $(x,y)$ coordinates in the plane of the
motion and measuring a new $\phi^\prime$ from the $x$ axis we may
write $\phi=\phi^\prime-\phi_0$ where $\phi_0$ is the azimuth of
$\hat{\bf e}$. With $\phi^\prime~{\bf v~h~r}~n~l$ all known it is then
possible to find a $\phi_0$ from the components of (\ref{8})/$e$. So
the implicit equation may be solved for $\hat{\bf e}$. There will be
very many solutions unless we restrict $\phi_0$ to be in the range
$-\pi/n$ to $+\pi/n$ and $\phi^\prime$ in the range $-\pi$ to
$+\pi$. I gave a discussion earlier$^6$ of the type of forces that
leave the magnitude of the eccentricity unchanged but slew its
direction.

\bigskip
\noindent{\it Approximating orbits in other potentials}
\medskip

Suppose we are given a potential $\psi (r)$ and an orbit within it is
defined by its pericentre at $r_p$ and its apocentre at $r_a$. Then
since $\dot{r}$ vanishes at these points the angular momentum of the
orbit is given by 
$$h^2=2\frac{\psi(r_p)-\psi (r_a)}{r_p^{-2}-r_a^{-2}}$$
and the energy of the orbit is given by
$$\varepsilon=\frac{h^2}{2r_p^2}-\psi(r_p)~.$$
The angle between perihelion and aphelion is then writing $r=u^{-1}$
$$\Phi=\int^{r_p^{-1}}_{r_a^{-1}} \left\{ 2h^{-2}\left[\varepsilon +
\psi(u^{-1})\right] -u^2\right\}^{-1/2}du~.$$ This may be compared with
the angle given by the orbit (\ref{5}) which is $\pi/n$.

Hence we may define the $n$ of the approximating orbit by
$n=\pi/\Phi$.

We shall make the angular momenta of the two orbits equal and define
the eccentricity by $$e=\frac{r_a-r_p}{r_a+r_p}$$ which is also in
conformity with (\ref{5}).

Evidently the $K$ of our approximating potential is already known via
$n$ since $K=h^2(1-n^2)$. To specify the approximating potential we
still need $\mu$ which we fix by making the perihelion distances
equal, which, via the eccentricity, implies the aphelion distances are
equal and hence
$$\mu=(h^2-K)(r^{-1}_p +r^{-1}_a)/2~.$$ Thus we have an approximation
scheme giving $n,~K,~h,~e,~\mu$ for any orbit specified by $r_p$ and
$r_a$ in any known potential $\psi(r)$. We expect the ${\bf e}$
defined by (\ref{8}) to be approximately constant for such orbits.

\bigskip
\noindent{\it The Eccentricity Vector for the Harmonic Oscillator}
\medskip

Newton$^7$ explained the connection between the Keplerian ellipse and
that generated by the two dimensional harmonic oscillator. Here we
follow Chandrasekhar's preferred path$^2$. Set $x+iy=z$ in the plane
of the orbit. The equation of the harmonic oscillator is then
$\frac{d^2z}{dt^2}+\omega^2z=0$ and its energy is
$E=\frac{1}{2}(\frac{d\overline{z}}{dt}\frac{dz}{dt} +
\omega^2\overline{z}z)$. Its angular momentum is
$h=\frac{1}{2i}(\overline{z}\frac{dz}{dt}-\frac{d\overline{z}}{dt}z)$. Now
consider the mapping of the complex plane $Z=z^2$. Following Newton we
ask whether the mapped path considered with a new time $\tau(t)$ can
be an orbit under a new central force. Evidently the angular momentum
of the $Z$ orbit is
$$\frac{1}{2i}\left(\overline{Z}\frac{dZ}{d\tau}-
\frac{d\overline{Z}}{d\tau}Z\right) = \frac{1}{i} |z|^2 \left(
\overline{z}\frac{dz}{d\tau}-\frac{d\overline{z}}{d\tau}z\right)
=2\frac{dt}{d\tau} |z|^2 h$$ so if the angular momenta are to be equal
then $\frac{d}{d\tau}=\frac{1}{2|z|^2} \frac{d}{dt}$. Chandrasekar (in
error!) omits the factor 2. Now using the $z$ equation of motion
$$\frac{d^2Z}{d\tau^2}= \frac{1}{4|z^2|} \frac{d}{dt}
\left(\frac{2}{\overline{z}} \frac{dz}{dt}\right) =
\frac{1}{2z\overline{z}^3} \left(\frac{d\overline{z}}{dt}\frac{dz}{dt}
-\omega^2\overline{z}z\right) = \frac{-\varepsilon}{z\overline{z}^3}
=-\frac{\varepsilon Z}{|Z|^3}$$ but the final expression shows us $Z$
is a motion under an inverse square law with a force constant
$\mu=GM=\varepsilon$ where $\varepsilon$ is the energy of the simple
harmonic orbit. Thus under the mapping the simple harmonic centred
ellipse becomes the Kepler eccentric ellipse. However the latter has a
conserved Hamilton eccentricity vector so what does that vector become
under the inverse transformation from Kepler's to Hooke's ellipse? In
the notation of this section vectors in the plane of the motion are
complex numbers. Since $d^2Z/d\tau^2=-\mu Z/|Z|^3$ we follow our well
trodden path and multiply by $h$ and integrate to find
$$h~dZ/d\tau=i\mu (Z/|Z|+e)$$
where $e$ is the (complex) constant of integration. To transform this
into the $z$ plane we write $d/d\tau=\frac{1}{2|z|^2}\frac{d}{dt},~Z=z^2$
and $\mu=\varepsilon$. So
\begin{equation} \label{9}
-\left(\frac{ih}{\varepsilon}
\frac{1}{\overline{z}}\frac{dz}{dt}+\frac{z}{\overline{z}}\right) =e~.
\end{equation}

I had to differentiate this extraordinary expression and show the
result to be zero before I believed that it was indeed a constant of
the motion! In the original $Z$ space $e$ pointed toward
perihelion. After the transformation $Z=z^2$, $e$ is still the same
complex number so is unchanged but now the perihelion in $z$ space
will be at half the angle to the real axis. This suggests that we
should be considering a new vector $\tilde{e}_p$ pointing to
perihelion and with the property that $e=\tilde{e}^2_p$. This
$\tilde{e}_p$ will then be a constant of the motion too and
furthermore the $\pm$ ambiguity in its definition reflects the fact
that the Hooke ellipse has two perihelia in opposite directions. There
is an intrinsic difficulty in the transformation that we have to place
a cut in the complex $Z$ plane to define $\sqrt{Z}$ properly. It we
place this cut arbitrarily then the direction of that cut intrudes
into the resultant formulae. It is much more sensible to take the cut
to be defined physically. Taking the cut along the real axis and that
toward perihelion along $e$ has advantages. Then both $e$ and
$\tilde{e}_p$ are real. Rewriting our equation to give us the velocity
we have
$$\frac{dz}{dt}=\frac{i\varepsilon}{h}(z+e\overline{z})$$
we may now rewrite this in vector form
$$\frac{d{\bf r}}{dt}
=\frac{\varepsilon}{h}\left\{\hat{\bf h}\times[{\bf
r}+\tilde{\bf e}_p .{\bf r}\tilde{\bf e}_p+\tilde{\bf e}_p \times
(\tilde{\bf e}_p\times {\bf r})]\right\}$$ which gives the velocity in
terms of the vector $\tilde{\bf e}_p$ that points to perihelion and
the position vector ${\bf r}$ in the orbital plane orthogonal to
$\hat{\bf h}$. To find the magnitude of $e$ and $\tilde{e}_p$ we
return to equation (\ref{9}). The general solution to the harmonic
equation is $z=pe^{i\omega t}+qe^{-i\omega t}$ where $p$ and $q$ are complex
numbers. In terms of $p$ and $q$,
$h=\omega(p\overline{p}-q\overline{q})$ and
$\varepsilon=\omega^2(p\overline{p}+q\overline{q})$. Putting these
expressions into equation (\ref{9}) we find
$$e=\frac{-2pq}{p\overline{p}+q\overline{q}}$$
choosing the real axis along $e$ means that $p=P e^{i\chi}$ and
$q=-Qe^{-i\chi}$ with $P$ and $Q$ real and positive, then
$z=(P-Q)\cos(\chi+\omega t)+i(P+Q)\sin(\chi+\omega t)$

$$e=\frac{2PQ}{P^2+Q^2}=\frac{a^2-b^2}{a^2+b^2}=\frac{e_\star^2}{2-e^2_\star}
~,$$ where $a,~b$ are the semi axes and $e_\star$ is the
`eccentricity' of the centred Hooke ellipse. Thus the magnitude of
$\tilde{e}_p$ is given by
$$\tilde{e}_p=\frac{e_\star}{\sqrt{2-e^2_\star}}~.$$

Of course $\tilde{\bf e}_p$ could be generalised to cases in which the
linear Hooke law is supplemented by an inverse cube repulsion, by
following the method given in this paper.

\vspace{2cm}

\centerline{\it References}
\medskip

\noindent(1) I. Newton, {\it Principia} (Royal Society, London)
Proposition 44, 1687.

\noindent(2) S. Chandrasekhar, {\it Newton's Principia for the Common
Reader}, (Clarendon Press, Oxford), pp. 184--191 \& p. 123, 1995.

\noindent(3) D. Lynden-Bell \& R. M. Lynden-Bell, {\it Notes \&
Recds. R. Soc.} {\bf 51}, 195, 1997.

\noindent(4) W. R. Hamilton, {\it Proc. R. Irish Acad.}, {\bf III},
Appendix p. 36, 1845.

\noindent(5) H. Goldstein, {\it Am. J. Phys.} {\bf 44}, 1123, 1976.

\noindent(6) D. Lynden-Bell, {\it The Observatory}, {\bf 120,
No. 1155}, 131, 2000.

\noindent(7) I. Newton, {\it Principia} (Royal Society, London) 2nd
Edition 1713 Proposition VII Corollary III, Proposition XI The same
otherwise. See (2) pp. 96 \& 119

\end{document}